\documentclass[a4paper,11pt]{article}
\pdfoutput=1 
\usepackage{jheppub} 

\usepackage[T1]{fontenc} 
\usepackage{caption} 
\usepackage{subcaption}

\usepackage[latin9]{inputenc}
\usepackage{float}
\usepackage{amsmath}
\usepackage{amssymb}
\usepackage{graphicx}
\usepackage{esint}
\usepackage{slashed}
 \usepackage{hyperref}
\usepackage{comment}
\usepackage{color}
\usepackage{microtype}
\usepackage{cleveref}
\usepackage{breakurl}
\usepackage{bbm}
\newcommand{\be}{\begin{equation}}
\newcommand{\ee}{\end{equation}}
\newcommand{\ben}{\begin{displaymath}}
\newcommand{\een}{\end{displaymath}}
\newcommand{\bea}{\begin{eqnarray}}
\newcommand{\eea}{\end{eqnarray}}

\def\be{\begin{equation}}
\def\ee{\end{equation}}
\def\bea{\begin{eqnarray}}
\def\eea{\end{eqnarray}}
\def\ba{\begin{array}}
\def\ea{\end{array}}
\def\bit{\begin{itemize}}
\def\eit{\end{itemize}}

 \makeatletter

\allowdisplaybreaks

\makeatother

\makeatletter
\DeclareRobustCommand{\rcite}[1]{%
  \rcite@aux#1,\@nil{#1}%
}
\def\rcite@aux#1,#2\@nil#3{%
  \if\relax#2\relax
    Ref.~\cite{#3}%
  \else
    Refs.~\cite{#3}%
  \fi
}
\makeatother

\hypersetup{
    colorlinks = true,
    citecolor = {blue},
    linkcolor = {blue},
    urlcolor = {blue},
}

 \title{\rm {\bf \huge  \boldmath Off-shell $N=2\to N=1$ reduction in 4D conformal supergravity }}

\author{Yusuke Yamada}
\affiliation{Stanford Institute for Theoretical Physics and Department of Physics, Stanford University, Stanford, CA 94305, USA}
  \emailAdd{yusukeyy@stanford.edu}


\notoc

\abstract{We discuss $N=2\to N=1$ reduction in four dimensional conformal supergravity. In particular, we keep the off-shell structure of supermultiplets (except hypermultiplets). As we will show, starting with (almost) off-shell conformal supergravity makes the procedure simpler than that from $N=2$ Poincar\'e supergravity, which makes it easier to show the correspondence to the standard $N=1$ conformal supergravity. We find that the $N=1$ superconformal symmetry is simply realized by truncating the gravitino multiplet. We also discuss the consistency with the original $N=2$ system and show the reduced $N=1$ conformal supergravity action.}

\begin{document}

\maketitle

 \newpage 
  \tableofcontents{}

 \parskip 6pt 

\section{Introduction}
$N=2$ supergravity appears as the effective field theory of superstrings compactified on a particular manifold, such as Calabi-Yau manifold. Such string compactification models can realize stable vacua thanks to supersymmetry. Realistic models in string theory may be constructed on such stable vacua. $N=2$ supersymmetry somehow needs to be broken at least to $N=1$ in realistic models having a chiral gauge symmetry as the standard model. In the context of string theory, it is achieved e.g. by introducing fluxes into compactified spaces, and realistic string models based on flux compactification have been intensively studied (see~\cite{Grana:2005jc,Douglas:2006es,Blumenhagen:2006ci} for review).

Partial breaking of supersymmetry $N=2\to N=1 $ may enable us to build more realistic models with chiral spectra, while keeping stability of vacua ensured by the $N=1$ supersymmetry. Spontaneous breaking from $N=2$ to $N=1$ had been thought to be impossible because of the no-go theorem for global~\cite{Witten:1981nf} and local supersymmetry~\cite{Cecotti:1984rk,Cecotti:1984wn}, but several ways to circumvent the theorem were proposed in~\cite{Hughes:1986fa,Hughes:1986dn,Bagger:1994vj,Antoniadis:1995vb} for global supersymmetry and in~\cite{Ferrara:1995gu,Ferrara:1995xi,Fre:1996js} for supergravity. Partial breaking models with non-Abelian gauge symmetry are discussed in~\cite{David:2003dh,Fujiwara:2004kc,Fujiwara:2005hj,Itoyama:2006ef,Maruyoshi:2006te}. In~\cite{Louis:2009xd}, it is shown that the spontaneous partial breaking condition in supergravity is systematically understood by using the embedding tensor formalism~\cite{deWit:2002vt,deWit:2005ub}. Corresponding low energy $N=1$ action is discussed in~\cite{Louis:2010ui}. In~\cite{Abe:2019svc}, an $N=2$ supergravity model interpolating partial breaking to full breaking is proposed.

Once $N=2$ supergravity is somehow broken to $N=1$, well below the energy scale of the breaking, we expect that the theory would be described in terms of $N=1$ supergravity. Such $N=1$ effective theories would be derived by truncating the gravitino multiplet of the second supersymmetry. The consistent $N=2\to N=1$ truncation was discussed in~\cite{Andrianopoli:2001zh,Andrianopoli:2001gm,DAuria:2005ipx}. 

In this paper, we perform the consistent reduction of $N=2$ conformal supergravity~\cite{deWit:1979dzm,deWit:1984rvr} to $N=1$ conformal supergravity~\cite{Kaku:1978nz,Kaku:1978ea,Kugo:1982cu}. Particularly, we keep the off-shell structure of the $N=2$ conformal supergravity except the hypermultiplet sector, which is originally an on-shell multiplet. In~\cite{Andrianopoli:2001zh,Andrianopoli:2001gm,DAuria:2005ipx}, the reduction was discussed by using on-shell Poincar\'e supergravity. It has been known that both $N=1$ and $N=2$ Poincar\'e supergravity can be viewed as conformal supergravity with particular gauge conditions. The presence of the extra gauge degrees of freedom is practically useful. Furthermore, we will keep the off-shell structure of the multiplets in $N=2$ conformal supergravity (except hypermultiplets, which are generally on-shell), and the off-shell formulation simplifies discussion significantly as one usually expects. We will explicitly show how the $N=2$ supermultiplets are decomposed into $N=1$ representations. As we will see, the truncation condition is simply understood as the absence of the ``$N=1$ gravitino multiplet''. We will show that, under the condition, the rest of the Weyl multiplet components, gauge and hypermultiplets, are precisely decomposed into the standard $N=1$ supermultiplets. The consistency with the ``parent'' $N=2$ theory leads to further conditions on matter multiplets, which reduces the number of $N=1$ multiplets as observed in~\cite{Andrianopoli:2001zh,Andrianopoli:2001gm,DAuria:2005ipx}.

The remaining part of this paper is organized as follows. In Sec.~\ref{sub}, we discuss the $N=1$ superconformal subgroup of the full $N=2$ group, and find how the gauge fields of $N=2$ superconformal symmetry form $N=1$ multiplets. In Sec.~\ref{red}, we impose the consistent truncation condition and discuss the decomposition of the rest of the fields into $N=1$ supermultiplets. We will see the complete agreement with the standard $N=1$ conformal supergravity. We rewrite the $N=2$ conformal action in terms of $N=1$ representations in Sec.~\ref{action}. We also discuss the consistency for the truncation condition derived in Sec.~\ref{sub} and find that half of the matter representations are required to be truncated. Finally, we conclude in Sec.~\ref{summary}. We show the $N=2$ and $N=1$ superconformal algebra and the transformation laws of representations in Appendix~\ref{A} and~\ref{B}, respectively.

Throughout this paper, we will use the notation of~\cite{Freedman:2012zz}.

\section{$N=1$ subgroup of $N=2$ superconformal group}\label{sub}
We discuss the $N=1$ superconformal subgroup of the full $N=2$ superconfromal group. The commutation relations and transformation laws of relevant multiplets in $N=2$ and $N=1$ superconformal algebra are shown in Appendix~\ref{A} and \ref{B}, respectively. 

We define the $N=1$ superconformal subgroup of $N=2$ superconformal group as follows. The subgroup consists of $\{ Q^1, Q_1, S^1,S_1, P_a, M_{ab}, K_a,D, T, U_1{}^1 \}$. Since $U_1{}^1$ is anti-Hermitian, we may parametrize $U_1{}^1={\rm i}T'$ with a Herimitian generator $T'$. The $\{Q,S\}$ commutator is given by (see \eqref{N2algebra})
\begin{equation}
\{Q_\alpha^1, S_1{}^{\beta}\}=-\frac12\delta_\alpha^\beta D-\frac14(\gamma^{ab})_\alpha^\beta M_{ab}+\frac{\rm i}2\delta_\alpha^\beta T-{\rm i}\delta_{\alpha}^\beta T'.
\end{equation}
The linear combination $T-2T'\equiv A$ becomes the chiral U(1) symmetry in the $N=1$ superconformal subgroup. Actually, since $[T',Q^1_\alpha]=-\frac {\rm i}{2} Q^1_\alpha$, we find $[A,Q^1_\alpha]=\frac{3{\rm i}}{2}Q^1_\alpha$, which reproduces the commutator $[A,Q]$ in the standard $N=1$ superconformal algebra~\eqref{N1algebra}. One can also check that $[A,S]$ commutator is correctly realized. We also define another U(1) symmetry, $T+T'=B$, which commutes with all generators in the $N=1$ superconformal group. The other commutation relations are intact. Thus, we find $N=1$ subgroup in $N=2$ superconformal algebra with an additional internal $U(1)_B$ symmetry.

 Next, let us see the transformation laws under the subgroup, particularly $Q$ and $S$ transformations:
 \begin{align}
\delta_{Q_1} e_{\mu}^a=&\frac12\bar{\epsilon}^1\gamma^a\psi_{\mu1}+\frac12\bar{\epsilon}_1\gamma^{a}\psi^{1}_\mu,\\
\delta_{Q_1}\psi_\mu^1=&\left(\partial_\mu+\frac{1}{2}b_\mu+\frac14\gamma^{ab}\omega_{\mu ab}-\frac {3{\rm i}}{2}\hat A_\mu\gamma^*\right)\epsilon^1-\gamma_\mu\eta^1,\label{gravtr}\\
\delta_{Q_1} b_\mu=&\frac12\bar{\epsilon}^1\phi_{\mu1}-\frac12\bar\eta^1\psi_{\mu 1}-\frac38\bar{\epsilon}^1\gamma_\mu\chi_1+{\rm h.c.},
\end{align}
where $\epsilon^1$ and $\eta^1$ are the spinorial transformation parameters for $Q_1$ and $S_1$, respectively.
Here $\hat A_\mu=\frac13(A_\mu+2{\rm i} V_\mu {}^1{}_1)$, which would be the gauge field for $U(1)_A$ of the $N=1$ subgroup. The transformation law of $\hat A_\mu$ is 
\begin{equation}
\delta_{Q_1} \hat A_\mu=-\frac {\rm i}{2}\bar{\epsilon}^1\phi_{\mu1}+\frac{\rm i}{8}\bar\epsilon^1\gamma_{\mu}\chi_1-\frac{\rm i}{2}\bar\eta^1\psi_{\mu1}+{\rm h.c.}.
\end{equation}
It is important to stress that the $N=2$ Weyl multiplet has extra ``matter'' components $T^{-}_{ab}$, $\chi^i$, $D$ in addition to gauge fields, which are necessary to close the algebra off-shell. However, in $N=1$ conformal supergravity, such extra fields are absent. One needs to remove the term with $\chi_1$ in order to reproduce the correct $N=1$ superconformal transformation. It is achieved by redefining $\phi_\mu^1$ as
\begin{equation}
\hat\phi_{\mu1}=\phi_{\mu1}-\frac14 \gamma_\mu \chi_1.
\end{equation}
Then, the transformation law becomes
\begin{equation}
\delta_{Q_1} \hat A_\mu=-\frac {\rm i}{2}\bar{\epsilon}^1\hat\phi_{\mu1}-\frac {\rm i}{2}\bar\eta^1\psi_{\mu1}+{\rm h.c.},
\end{equation}
which is the correct transformation law of $\hat A_\mu$ if we identify $\epsilon^1=P_L\epsilon$, $\eta^1=P_R\eta$ and $\hat\phi_{\mu1}=P_L\phi_\mu$, where $\epsilon$ and $\eta$ are $N=1$ $Q$- and $S$-transformation parameter, respectively, and $\phi_\mu$ is $S$-gauge field in $N=1$ conformal supergravity. One also needs to deform the transformation of $b_\mu$ as follows. The transformation law of $b_\mu$ is now
\begin{equation}
\delta_{Q_1} b_\mu=\frac12\bar{\epsilon}^1\hat\phi_{\mu1}-\frac12\bar\eta^1\psi_{\mu 1}-\frac14\bar{\epsilon}^1\gamma_\mu\chi_1+{\rm h.c.}
\end{equation}
To eliminate the term with $\chi_1$, we use $\epsilon$ dependent $K$-transformation, $\lambda_{K\mu}=\frac18\bar{\epsilon}^1\gamma_\mu\chi_1+{\rm h.c.}$. Note that this modification does not affect transformation laws of other gauge fields since only $b_\mu$ transforms as $\delta_Kb_\mu=2\lambda_{K\mu}$ under~$K$ and other independent fields are inert under $K$.
Thus, the $N=1$ transformation is realized as $\hat\delta_Q=\delta_{Q_1}(\epsilon^1)+\delta_K(\frac18\bar{\epsilon}^1\gamma_\mu\chi_1)$, and $\hat\delta_S=\delta_S(\eta^1)$.

The transformation law of $B_\mu=\frac12(A_\mu+{\rm i}V_{\mu}{}_1{}^1)$ is simply given by
\begin{equation}
\hat\delta B_\mu=-\frac{3{\rm i}}{8}\bar{\epsilon}^1\gamma_\mu\chi_1+{\rm h.c.}.
\end{equation}
We also note that the transformation law of $\chi^1$ and of $D$ are 
\begin{equation}
\hat\delta\chi^1=\frac12 D \epsilon^1+\frac {\rm i}{6}\gamma^{ab}({\rm i}\hat{R}_{ab}(U_1{}^1)+\hat{R}_{ab}(T))\epsilon^1,
\end{equation} 
and
\begin{equation}
\hat\delta D=\frac12\bar\epsilon^1\slashed{D}\chi_1+{\rm h.c.}.
\end{equation}
One finds that the curvature combination ${\rm i}\hat{R}_{ab}(U_1{}^1)+\hat{R}_{ab}(T)$ depends only on $B_\mu$ (and its covariatization terms), and therefore, $(B_\mu, \chi^1, D)$ form a vector multiplet in $N=1$ subgroup. 

The rest of components in the $N=2$ Weyl multiplet transforms as
\begin{align}
\hat\delta\psi^2_\mu=&V_\mu{}^2{}_1\epsilon^1+\frac{1}{16}\gamma^{ab}T_{ab}^{-}\gamma_\mu\epsilon_1,\\
\hat\delta V_{\mu1}{}^2=&-\bar{\epsilon}_1\phi_\mu^2+\frac34\bar\epsilon_1\gamma_\mu\chi^2-\bar{\eta}_1\psi_\mu^2,\\
\hat\delta T_{ab}^{-}=&2\bar{\epsilon}^1\hat{R}_{ab}(Q^2),\\
\hat\delta\chi^2=&\frac16\gamma^{ab}\left[\frac14\slashed{D}T_{ab}^-\epsilon_1-\hat{R}_{ab}(U_1{}^2)\epsilon^1-\frac12T_{ab}^-\eta_1\right].
\end{align}
The gauge field $\phi^2_\mu$ is a composite field given by
\begin{equation}
\phi^2_\mu=-\frac12\gamma^\nu R_{\mu\nu}'(Q^2)+\frac{1}{12}\gamma_\mu\gamma^{ab}R_{ab}'(Q^2)+\frac14\gamma_\mu\chi^2,
\end{equation}
where
\begin{equation}
R_{\mu\nu}'(Q^2)=2\left(\partial_{[\mu}+\frac12b_{[\mu}+\frac14\gamma^{ab}\omega_{[\mu ab}+\frac{\rm i}{2}\hat{A}_{[\mu}-\frac{4\rm i}{3}B_{[\mu} \right)\psi^2_{\nu]}+2V_{[\mu}{}^2{}_1\psi^1_{\nu]}.
\end{equation}
Except the covariantization terms, $(\psi_\mu^2, V_{\mu1}{}^2,T_{ab}^-,\chi^2)$ transform to each other under the $N=1$ subgroup, and they seem to form an $N=1$ gravitino multiplet.

In summary of this section, we have identified $N=1$ superconformal subgroup of $N=2$ superconformal group, and we find that the $N=2$ Weyl multiplet can be decomposed into $N=1$ Weyl, gauge (vector), and spin-$3/2$ gravitino multiplet. Note that, however, these multiplets cannot be the standard $N=1$ superconformal multiplets because of the following reason: The ``Weyl multiplet'' in the $N=1$ subgroup is gauged under the second supersymmetry, whereas the standard $N=1$ Weyl multiplet is singlet under any symmetries but $N=1$ superconformal symmetry. Therefore, even though the subgroup structure looks similar to the standard $N=1$ conformal supergravity, the system is different and cannot be expressed in terms of the standard $N=1$ conformal supergravity. However, as we will see in the next section, the $N=1$ subgroup becomes the standard one once we truncate the second supersymmetry and its gauge field $\psi_\mu^2$ in a simple but consistent way.

\section{Reduction to $N=1$ conformal supergravity} \label{red}
In the previous section, we have discussed the structure of $N=1$ superconformal subalgebra. Here, we will consider the truncation to the standard $N=1$ superconformal system by eliminating gauge fields of extra symmetries, such as the second supersymmetry, and its superpartners under the subalgebra.

As we have shown in the previous section, there is a gravitino multiplet formed by $(\psi_\mu^2, V_{\mu1}{}^2,T_{ab}^-,\chi^2)$ in $N=1$ subgroup. If we set 
\begin{equation}
\psi^2_\mu=\epsilon_2=0,\label{psi2}
\end{equation}
the consistency of the remaining $N=1$ superconformal transformation requires all other superpartners to vanish, namely, 
\begin{equation}
V_{\mu1}{}^2=T_{ab}^-=\chi^2=0.\label{psi2sp}
\end{equation}
No further condition is required from the consistency with the remaining $N=1$ superconformal symmetry. These conditions also lead to $\phi^2_\mu=0$, which means the second $S$-supersymmetry generated by $S^2$ and $S_2$ is absent and we should set $\eta^2=0=\eta_2$. The $N=1$ transformation of the gravitino multiplet vanishes under \eqref{psi2} and \eqref{psi2sp}. We have not yet discussed the consistency of \eqref{psi2} and \eqref{psi2sp} with the $N=2$ superconformal action and their equations of motion. In particular, $N=2$ Lagrangian has terms linear in gravitino multiplet components, which are generally non-vanishing in their equations of motion even if we impose the truncation conditions \eqref{psi2}, \eqref{psi2sp}. Therefore, additional constraints should be imposed for consistency with the original $N=2$ action. We will discuss such conditions in Sec.~\ref{auxconst}. In the following, we just assume  \eqref{psi2} and \eqref{psi2sp}.

The set of gauge fields for $N=1$ superconformal subgroup transforms consistently with the standard $N=1$ superconformal transformation as discussed in the previous section. Let us look at the transformation of the $N=1$ gauge multiplet $(B_\mu, \chi^1, D)$, which originates from the $N=2$ Weyl multiplet. In order to fix normalization factors, we define 
\begin{equation}
P_L\hat\chi=-\frac{3{\rm i}}{4}\chi^1,
\end{equation} 
and then, we find
\begin{equation}
\delta_{ N=1} B_\mu=-\frac12\bar\epsilon\gamma_\mu\hat\chi,
\end{equation}
where $\hat\chi$ is a Majorana spinor, and we have identified $N=1$ supersymmetry transformation parameter $\epsilon$ as $P_L\epsilon=\epsilon^1$ (equivalently $\epsilon_1=P_R\epsilon$). Here and hereafter, $\delta_{N=1}$ denotes the $\hat\delta$-transformation with the truncation conditions \eqref{psi2}, \eqref{psi2sp}.
The combination of the curvature ${\rm i}\hat{R}_{ab}(U_1{}^1)+\hat{R}_{ab}(T)$ becomes
\begin{equation}
{\rm i}\hat{R}_{ab}(U_1{}^1)+\hat{R}_{ab}(T)=4\partial_{[\mu}B_{\nu]}+2\bar\psi_{[\mu}\gamma_{\nu]}\hat\chi\equiv 2\hat{F}_{\mu\nu}(B).
\end{equation}
The field strength $\hat F_{\mu\nu}(B)$ is the covariant field strength of the gauge vector $B_\mu$.
Here, we have defined the $N=1$ gravitino $\psi_\mu$ as
\begin{equation}
P_L\psi_\mu=\psi^1_\mu,
\end{equation}
which is consistent with the identification $\epsilon^1=P_L\epsilon$.
The transformation laws of $\hat \chi$ and $\hat D\equiv-\frac{3}{4}D$ are
\begin{equation}
\delta_{ N=1} \hat\chi=\frac {\rm i}2\hat D\gamma_*\epsilon+\frac14\gamma^{ab}\hat F_{ab}(B)\epsilon,
\end{equation}
\begin{equation}
\delta_{ N=1}\hat{D}=\frac{\rm i}{2}\bar\epsilon\gamma_*\gamma^\mu D_\mu\hat\chi.
\end{equation}
The set of the transformation laws of $(B_\mu,\hat\chi, \hat D)$ correctly realizes that of an $N=1$ gauge multiplet, as we expected (see \eqref{N1gauge}, \eqref{N1gaugino}, \eqref{N1D}). This is a nontrivial check for the correct truncation.

The second nontrivial check is the curvature of the new chiral U(1) gauge field $\hat A _\mu=\frac13(A_\mu-2{\rm i}V_{\mu1}{}^1)$, which is given by
\begin{equation}
\hat{R}(A)\equiv\frac{1}{3}(\hat R_{\mu\nu}(T)-2{\rm i} R_{\mu\nu}(U_1{}^1))=2\partial_{[\mu}\hat{A}_{\nu]}-{\rm i}\bar\psi^1_{[\nu}\hat\phi_{\mu]1}+{\rm i}\bar\psi_{[\nu1}\hat\phi_{\mu]}^1=2\partial_{[\mu}\hat{A}_{\nu]}+{\rm i}\bar\psi_{[\mu}\gamma_*\phi_{\nu]},
\end{equation}
where we have identified the S-supersymmetry gauge field $\phi_\mu$ as $P_L\phi_\mu=\hat\phi_1$. This combination precisely reproduces the curvature of the chiral $U(1)_A$ symmetry in the standard $N=1$ superconformal group. Note that, $\hat{\phi}^1_\mu$ is not the one of original $N=2$ $S$-supersymmetry, $\phi^1_\mu$, but the shifted one, $\hat \phi^1_\mu=\phi^1_\mu-\frac14 \gamma_\mu \chi^1$, and the $\chi^1$ dependence in the curvature is completely absorbed into $\phi_\mu$.  One can check that the same replacement $\phi^1_\mu\to\hat\phi^1_\mu$ leads to correct curvatures of other generators. Note also that, special conformal boost gauge field $f_\mu{}^a$ also needs to be shifted as
\begin{equation}
f_\mu^a\to \hat f_\mu{}^a=f_\mu{}^a+\frac18e_\mu{}^aD.
\end{equation}
This modified gauge field correctly follows the standard $N=1$ superconformal algebra.\footnote{This shift is also important to make the $N=2$ curvature constraints that of $N=1$. Taking into account the shifts of $\phi^1_\mu$ and $f_\mu{}^a$ in addition to the truncation conditions, we find that the $N=2$ curvature constraints precisely reproduce that of the standard $N=1$ conformal supergravity.}
Thus, our simple truncation gives the standard $N=1$ Weyl multiplet and an additional gauge multiplet $(B_\mu,\hat\chi, \hat D)$ of an internal $U(1)_B$ symmetry.

\subsection{$N=1$ truncation for matter sector}
Here, we discuss the $N=1$ reduction of vector and hyper-multiplets. As we show below, both the vector and the hyper-multiplets are simply decomposed into $N=1$ multiplets without truncating any components.

\subsubsection{Vector multiplets}

 First, we consider a vector multiplet~\cite{deWit:1984rvr} made of $(X^I, \Omega^I_i, Y_{ij}^I, A_{\mu}^I)$, where $I$ is an index for vector multiplets $I=1,\cdots n_V$, and $i,j$ denote $SU(2)$ R-symmetry indices. $X^I$ is a complex scalar, $Y_{ij}^I$ is an SU(2) triplet symmetric tensor $Y_{ij}^I=Y_{ji}^I$, $\Omega^I_i=P_L\Omega^I_i$ is a Weyl spinor, and $A_\mu^I$ is a gauge vector of (non-)Abelian group. The $N=1$ part of the supersymmetry and $S$-supersymmetry transformation is given by
\begin{align}
\delta_{N=1} X^I=&\frac12\bar{\epsilon}^1\Omega_1^I,\\
\delta_{N=1}\Omega_1^I=&\slashed{D}X^I\epsilon_1+Y_{11}^I\epsilon^1+2X^I\eta_1,\\
\delta_{N=1} Y_{11}^I=&\frac12\bar\epsilon_1\slashed{D}\Omega_1^I- f_{JK}^I\bar\epsilon_1X^J\Omega^{2K},\\
\delta_{N=1} A^I_\mu=&\frac12\bar\epsilon_1\gamma_\mu\Omega_2^I+{\rm h.c.},\\
\delta_{N=1}\Omega_2^I=&-\frac14\gamma^{ab}F^I_{ab}\epsilon^1+Y_{12}^I\epsilon^1-X^J\bar{X}^Kf_{JK}^I\epsilon^1,\\
\delta_{N=1} Y^I_{12}=&\frac14\bar\epsilon_1\slashed{D}\Omega_2^I+\frac12f_{JK}^I\bar\epsilon_1X^J\Omega^{1K}-{\rm h.c.},
\end{align}
where $f_{JK}^I$ is a structure constant of the algebra, which the gauge field obey.
It seems that $(X^I,\Omega^I_1, Y_{11}^I)$ form an $N=1$ chiral multiplet and $(A_\mu^I,\Omega_2^I, Y_{12}^I)$ form an $N=1$ gauge (vector) multiplet. To make the $N=1$ structure more clear, we need to redefine the auxiliary component $Y_{12}^I$ as $\hat{Y}^I_{12}=Y_{12}^I-f_{JK}^IX^{J}\bar{X}^{\bar K}$. Then, the transformation law of $\Omega_2^I$ and $\hat{Y}_{12}$ are given by
\begin{align}
\delta_{N=1}\Omega_2^I=&-\frac14\gamma^{ab}F^I_{ab}\epsilon^1+\hat{Y}_{12}^I\epsilon^1\\
\delta_{N=1} \hat{Y}^I_{12}=&\frac14\bar\epsilon_1\slashed{D}\Omega_2^I.
\end{align}
Also, we need to notice that the scalar $X^I$ is gauged under the internal symmetry and transforms as the adjoint representation $\delta X^I=k^I_J\theta^J=X^Kf^I_{KJ}\theta^J$. Then, the transformation laws of $(X^I,\Omega^I_1, Y_{11})$ are consistent with that of $N=1$ superconformal chiral multiplet.

Let us give the standard normalization to those $N=1$ multiplets. The chiral multiplets consist of $(X^I, P_L\chi^I\equiv\frac{1}{\sqrt{2}}\Omega_1^I,{\bf F}^I\equiv Y_{11}^I)$, and their transformation laws are given by
\begin{align}
\delta_{ N=1} X^I=&\frac{1}{\sqrt2}\bar\epsilon P_L\chi^I,\\
\delta_{ N=1}P_L\chi^I=&\frac{1}{\sqrt{2}}P_L(\slashed{D}X^I+{\bf F}^I)\epsilon,\\
\delta_{ N=1}{\bf F}^I=&\frac{1}{\sqrt2}\bar\epsilon\slashed{D}P_L\chi^I+f_{JK}^I\bar\epsilon X^JP_R\lambda^K,
\end{align}
which are the standard $N=1$ transformations of a chiral multiplet (see \eqref{N1CS}, \eqref{N1CW}, \eqref{N1F}). Here we have defined the gaugino $\lambda^I\equiv -\Omega^{2I}-\Omega_{2}^I$. In $N=1$ conformal supergravity, there are two important quantum numbers characterizing multiplets, the Weyl and the chiral weight $(w,n)$. Also, in our case, there is an additional $U(1)_B$ gauge symmetry originating from $N=2$ superconformal symmetry. In the following, we use $n_B$ as the U(1) charge. From the normalization of vectors $(\hat A_\mu, B_\mu)$, we find the charges of the chiral multiplet $X^I$ to be $(w,n,n_B)=(1,1,4/3)$.

The $N=1$ vector multiplet consists of $(A_\mu^I, \lambda^I=-\Omega_2^I-\Omega^{2I}, {\bf D}^I\equiv 2{\rm i}\hat{Y}_{12}^{I})$, and their $Q$-transformations are
\begin{align}
\delta_{N=1}A_\mu^I=&-\frac12\bar\epsilon\gamma_\mu\lambda^I,\\
\delta_{N=1}\lambda^I=&\frac14\gamma^{ab}\hat{F}_{ab}^I+\frac {\rm i}2\gamma_*{\bf D}^I\epsilon,\\
\delta_{N=1}{\bf D}^I=&\frac{\rm i}{2}\bar\epsilon\gamma_*\slashed{D}\lambda^I.
\end{align}
All of them are $S$-inert as the standard $N=1$ gauge multiplet (see \eqref{N1gauge}, \eqref{N1gaugino}, \eqref{N1D}). This gauge multiplet has weights $(w,n,n_B)=(0,0,0)$. As we have shown, the $N=2$ vector multiplet can be decomposed into $N=1$ chiral and gauge multiplets under the conditions~\eqref{psi2} and \eqref{psi2sp}.

\subsubsection{Hypermultiplets}\label{hyperreduction}

Next, let us consider the hypermultiplet sector~\cite{deWit:1984rvr,deWit:1999fp} which consists of $(q^X, \zeta^{\cal A} \ (\zeta_{{\cal A}}))$, where $X$ is the target space index and runs over $X=1,\cdots, 4n_H$, $\mathcal A$ is the tangent space one, $\mathcal{A}=1,\cdots, 2n_H$, and $n_H$ is the number of hypermultiplets. The real scalars $q^X$ are the coordinate of hyper-K\"ahler manifold, and $\zeta^{\mathcal A}$ ($\zeta_{\mathcal A}$) are left (right)-handed Weyl spinors. The $\hat\delta$-transformations are given as follows:
\begin{align}
\delta_{N=1} q^X=&-{\rm i}\bar\epsilon^{1}\zeta^{\cal A}f^X{}_{1\cal A}+{\rm i}\rho^{\bar{\cal A}{\cal B}}\bar\epsilon_1\zeta_{\bar{\cal A}}f^X{}_{2\mathcal B},\\
\delta_{N=1}\zeta^{\cal A}=&\frac{\rm i}{2}f^{1\mathcal A}{}_X\slashed{D}q^X\epsilon_1-\zeta^{\cal B}\omega_{X\mathcal B}{}^{\cal A}\delta_{N=1} q^X-{\rm i}\bar{X}^Ik_{I}^Xf^{2\mathcal{A}}{}_X\epsilon^1+{\rm i}f^{1\mathcal{A}}{}_Xk_D^X\eta_1,\label{ztr1}\\
\delta_{N=1} \zeta_{\bar{\cal A}}=&\frac{\rm i}{2}f^{2\mathcal B}{}_X\rho_{\mathcal B{\bar{\cal A}}}\slashed{D}q^X\epsilon^1-\zeta_{\bar{\cal B}}\bar{\omega}_{X}{}^{\bar{\mathcal B}}{}_{\bar{\cal A}}\delta_{N=1} q^X+{\rm i}X^Ik_I^Xf^{1\mathcal B}{}_X\rho_{\mathcal{B}{\bar{\cal A}}}\epsilon_1+{\rm i}f^{2\mathcal{B}}{}_X\rho_{{\cal B}\bar{\cal A}}k_D^X\eta^1,
\end{align}
where $\rho_{\mathcal A\bar{\cal B}}$ is a covariantly constant tensor satisfying $\rho_{\mathcal A\bar{\cal B}}\rho^{\bar{\cal B}\cal C}=-\delta_{\cal A}^{\cal C}$ and $\rho^{\bar{\cal A}\cal B}=\left(\rho_{\mathcal A\bar{\cal B}}\right)^*$ and $\omega_{X\cal A}{}^{\cal B}(q)$ is the connection for the tangent space reparametrization. $f^X{}_{i\mathcal A}(q)$ and its inverse $f^{i\mathcal {A}}{}_X(q)$ are frame fields connecting $\mathcal A$ and $X$ indices, which satisfy the following relations
\begin{equation} 
f^{i\mathcal A}{}_Yf^X{}_{i\mathcal A}=\delta^X_Y, \qquad f^{i\mathcal A}{}_Xf^X{}_{j\mathcal B}=\delta^i_j\delta_{\cal B}^{\cal A}.
\end{equation}
There is also a reality condition
\begin{equation}
\left(f^{i\mathcal{A}}{}_X\right)^*=f^{j\mathcal{B}}{}_X\varepsilon_{ji}\rho_{\mathcal B \bar{\cal A}},\label{reality}
\end{equation}
where $\varepsilon_{ij}$ is an anti-symmetric tensor and we take $\varepsilon_{12}=1$.
Using these relations, one finds
\begin{equation}
f^{1\cal A}{}_X\delta_{N=1} q^X=-{\rm i}\bar\epsilon^{1}\zeta^{\cal A}, \qquad f^{2\mathcal B}{}_X\delta_{N=1} q^X={\rm i}\rho^{{\cal A}{\cal B}}\bar\epsilon_1\zeta_{{\cal A}}.
\end{equation}
The vectors $k_D^X, k_I^X$ are defined by the $D$, $U_i{}^j$ and the internal gauge transformations of the hyperscalar $q^X$,
\begin{equation}
\delta q^X=\lambda_D k_D^X+\lambda_i{}^jf^{i\mathcal A}{}_Yf^X{}_{j\mathcal A}k_D^Y+\theta^Ik_I^X,\label{htr}
\end{equation} 
where $\lambda_D$, $\lambda_i{}^j$ and $\theta^I$ are the transformation parameters of $D$, $U_i{}^j$ and internal gauge symmetry $T_I$, respectively. See~\cite{deWit:1999fp} for more details of the superconformal hypermultiplet.

The transformation law of $\zeta^{\mathcal A}$ is not covariant under the reparametrization $\zeta^{\mathcal A} \to \zeta^{\cal B}U_{\cal B}{}^{\cal A}(q)$. The second term in \eqref{ztr1} originates from such non-covariance. Therefore, we define the covariant supersymmetry transformation
\begin{equation}
\delta_{N=1}^{\rm cov}\zeta^{\cal A}=\delta_{N=1} \zeta^{\cal A}+\zeta^{\cal B}\omega_{X\mathcal B}{}^{\cal A}\delta_{N=1} q^X.
\end{equation}
For more details of the covariant formulation, see \cite{Freedman:2016qnq,Freedman:2017obq}. The covariant version of the transformation is given by
\begin{equation}
\delta_{N=1}^{\rm cov}\zeta^{\cal A}=\frac{\rm i}{2}f^{1\mathcal A}{}_X\slashed{D}q^X\epsilon_1-{\rm i}\bar{X}^Ik_{I}^Xf^{2\mathcal{A}}{}_X\epsilon^1+{\rm i}f^{1\mathcal{A}}{}_Xk_D^X\eta_1.
\end{equation}
Let us introduce the following sections
\begin{equation}
A^{i\mathcal A}\equiv f^{i\cal A}{}_X k_D^X.
\end{equation}
The covariant supersymmetry transformation $\hat\delta_{\rm cov}A^{i\mathcal A}=\hat\delta A^{i\mathcal A} +A^{i\mathcal B}\omega_{X\mathcal B}{}^{\cal A}\hat\delta q^X$ is given by
\begin{equation}
\delta_{N=1}^{\rm cov}A^{1\mathcal A}=-{\rm i}\bar\epsilon^{1}\zeta^{\cal A},\qquad \delta_{N=1}^{\rm cov}A^{2\mathcal B}={\rm i}\rho^{\bar{\cal A}{\cal B}}\bar\epsilon_1\zeta_{\bar{\cal A}},
\end{equation}
where we have used the property of the closed homothetic Killing vector $\nabla_Y k_{D}^X=\delta_Y^{X}$ and $\nabla_X$ is the target space covariant derivative. These transformations correspond to that of $N=1$ chiral and anti-chiral superfields, respectively. In the following, we call $A^{i\mathcal A}$ as
\begin{equation}
\Phi^{\mathcal A}=A^{1\mathcal A}, \qquad \bar\Phi_{\bar{\cal A}}=-\rho_{\mathcal B \bar{\cal A}}A^{2\mathcal B}.\label{ftr}
\end{equation}
We note that $\left(\Phi^{\cal A}\right)^*=\bar\Phi_{\bar{\mathcal A}}$, which follows from the reality condition \eqref{reality}. We rewrite the $N=1$ transformations laws \eqref{ftr} as 
\begin{equation}
\delta_{N=1}^{\rm cov}\Phi^{\cal A}=\frac{1}{\sqrt2}\bar\epsilon P_L\hat\zeta^{\cal A},\qquad \delta_{N=1}^{\rm cov}\bar\Phi_{\bar{\cal A}}=\frac{1}{\sqrt2}\bar\epsilon P_R\hat\zeta_{\bar{\mathcal A}},
\end{equation}
where $\hat\zeta^{\mathcal A}=-\sqrt{2}{\rm i}\zeta^{\cal A}$ and $\hat\zeta_{\bar{\mathcal A}}=\sqrt{2}{\rm i}\zeta_{\bar{\cal A}}$.
The covariant transformation of $\hat\zeta^{\cal A}$ and $\hat\zeta_{\bar{\cal A}}$ are 
\begin{align}
\delta^{\rm cov}_{N=1}\hat\zeta^{\cal A}=&\frac{1}{\sqrt2}P_L\slashed{\nabla}\Phi^{\mathcal A}\epsilon-\sqrt{2}\bar{X}^I\bar{t}_{I}{}^{\bar{\cal B}\bar{\cal C}}\bar{\Phi}_{\bar{\cal B}}(d^{-1})_{\bar{\cal C}}^{\cal A}P_L\epsilon+\sqrt2\Phi^{\mathcal A}P_L\eta,\label{ztr}\\
\delta^{\rm cov}_{N=1} \hat{\zeta}_{\bar{\cal A}}=&\frac{1}{\sqrt 2}P_R\slashed{\nabla}\bar{\Phi}_{\bar{\cal A}}\epsilon-\sqrt2X^It_{I{\cal BC}}\Phi^{\cal B}(d^{-1})^{\cal C}_{\bar{\cal A}}P_R\epsilon+\sqrt2\bar\Phi_{\bar{\cal A}}P_R\eta,
\end{align}
where we have defined the covariant derivative of $\nabla_\mu\Phi^{\mathcal A}$ as
\begin{equation}
\nabla_\mu\Phi^{\mathcal A}\equiv D_{\mu}q^X\nabla_{X}\Phi^{\mathcal A}=D_\mu q^Xf^{1\cal A}{}_X.
\end{equation}
Here we have used the following facts in deriving the expression~\eqref{ztr}: The frame fields are covariantly flat, $\nabla_Yf^{i\cal A}{}_X=0$. $\nabla_Y k_{D}^X=\delta_Y^{X}$ and $f^{1\cal A}{}_X=\nabla_X(k_D^Yf^{1\cal A}{}_Y)=\nabla_X\Phi^{\cal A}$. From the definition of $\Phi^{\cal A}$, we find $k_D^X=f^X{}_{1\cal A}\Phi^{\cal A}$. Also, the commutativity of the internal symmetry and the dilatation leads to $k_{I}^X=k_D^Y\nabla_Yk_I^X$, and we find $k_I^X=\Phi^{\cal C}f^Y{}_{1\cal C}\nabla_Yk_I^X$ and $\nabla_Y\nabla_Xk_D^Z=0$. Using these facts, one finds
\begin{equation}
-{\rm i}X^Ik_I^Xf^{1\cal B}{}_X\rho_{{\cal B}\bar{\cal A}}=-{\rm i}X^It_{I\cal CB} \Phi^{\cal B}( d^{-1})^{\cal C}_{\bar{\cal A}},
\end{equation}
where $\tilde{t}_{I\cal AB}\equiv C_{\cal AC}f^Y{}_{1\cal B}\nabla_Yk_I^Xf^{1\cal C}{}_{X}$ is a covariantly constant symmetric tensor $\tilde{t}_{I\cal AB}=\tilde{t}_{I\cal BA}$, $\bar{t}_{I}^{\bar{\cal A}\bar{\cal B}}=(t_{I\cal AB})^*$ and $(d^{-1})^{\cal A}_{\bar{\cal B}}$ is the inverse of $d^{\bar {\cal B}}_{\cal A}$. 

Since the hypermultiplets are on-shell multiplets, the corresponding $N=1$ superfields are also on-shell multiplets and do not have auxiliary fields. This is why there is no $F$-term, which appears in the standard $N=1$ off-shell chiral multiplets' transformation. We can read off the on-shell value of F-term of $\Phi^{\cal A}$ from the transformation of $\hat\zeta^{\cal A}$. Note that the transformation of Weyl spinor has the term $\frac{1}{\sqrt2}{\bf F}P_L\epsilon$ for an off-shell chiral multiplet. Thus, the transformation law of $\hat \zeta^{\cal A}$ reads the on-shell F-term of $\Phi^{\cal A}$,
 \begin{equation}
{\bf F}^{\mathcal A}=-2\bar{X}^I\bar{t}_{I}{}^{\bar{\cal B}\bar{\cal C}}\bar{\Phi}_{\bar{\cal B}}(d^{-1})_{\bar{\cal C}}^{\cal A}.\label{onshellF}
\end{equation}
We will show that this value is consistent with the action of hypermultiplets. One may wonder why there is no fermionic term in the on-shell F-term, which usually exists (see \cite{Freedman:2012zz}, for example). In \cite{Freedman:2016qnq,Freedman:2017obq} it is shown that the covariant F-term on shell is given by a purely bosonic term. Therefore our covariant chiral multiplet $\Phi^{\cal A}$ is consistent with the results of~\cite{Freedman:2016qnq,Freedman:2017obq}. 

It is also worth noting that, to realize $N=1$ superconformal symmetry, there is no need to reduce the degrees of freedom of hypermultiplets. The transformation laws of the chiral multiplet $(\Phi^{\mathcal A}, \hat\zeta^{A})$ and the anti-chiral one $(\bar\Phi_{\bar{\mathcal A}},\hat \zeta_{\bar{\cal A}})$ are precisely that of the (covariantly modified) standard ones. 

Finally, let us discuss the charges $(w,n,n_B)$ of the chiral multiplets. The dilatation and SU(2) transformation of $A^{i{\cal A}}$ following from \eqref{htr} is
\begin{equation}
\delta _{D,SU(2)}A^{i\cal A}=\lambda_DA^{i\cal A}+\lambda_j{}^iA^{j{\cal A}}.
\end{equation}
From this expression, one can read off the weights of $\Phi^{\cal A}$ to be $(w,n,n_B)=(1,1,-\frac{2}{3})$ (and of $\bar\Phi_{\bar{\cal A}}$ to be $(1,-1,\frac{2}{3})$).\footnote{One of the ways to confirm the charge assignments is to check how the gauge fields are coupled in the covariant derivative.}

\section{Reduction of superconformal action from $N=2$ to $N=1$}\label{action}
In this section, we discuss whether the $N=2$ superconformal action is consistently reduced to that of the standard $N=1$ conformal supergravity. The standard $N=1$ superconformal action of chiral and gauge multiplets is
\begin{equation}
S^{N=1}=[{\cal N}(\Phi^I,\bar\Phi^{\bar J})]_D+[W(\Phi^I)]_F+[f_{AB}(\Phi^I)\bar\lambda^AP_L\lambda^B]_F,
\end{equation}
where ${\cal N}$ is a real function of chiral multiplets $\Phi^I$ and its conjugate, $W$ and $f_{AB}$ are holomorphic function of chiral multiplets, and $\lambda^A$ is a gaugino. $[\cdots]_{D,F}$ denotes the superconformal $D$- and $F$-term density formulae, respectively~\cite{Kugo:1982cu}. The bosonic part of the action is given by
\begin{align}
S^{N=1}|_B=&\int d^4xe\Biggl[-\frac{\cal N}{6}R-{\rm i}{\cal N}_Ik_A^ID^A-{\cal N}_{I\bar{J}}(D_\mu X^ID^\mu\bar X^{\bar J}-F^I\bar{F}^{\bar{J}})+(W_IF^I+{\rm h.c.})\nonumber\\
&-\frac{1}{4}({\rm Re}f_{AB})F^{A}_{\mu\nu}F^{B\mu\nu}+\frac{\rm i}{4}({\rm Im}f_{AB})F^A_{\mu\nu}\tilde{F}^{B\mu\nu}+\frac12 ({\rm Re}f_{AB})D^AD^B\Biggr],
\end{align}
where $e={\rm det}(e^a_\mu)$, $R$ is the Ricci scalar, ${\cal N}_{I}=\partial_I{\cal N}$, ${\cal N}_{I\bar{J}}=\partial_I\partial_{\bar{J}}N$, $k_A^I$ is the gauge transformation of $\Phi^I$, $\delta\Phi^I=\theta^Ak_{A}^I$ and $\theta^A$ is the gauge transformation parameter. The complex and real scalar $F^I$ and $D^A$ are F- and D-term of chiral and gauge multiplets, respectively.

Let us first discuss the $N=2$ vector multiplet action under the condition \eqref{psi2sp}, which has the following bosonic part,
\begin{align}
S_{\rm vec}|_B=\int d^4x e\Biggl[&-\frac{1}{6}NR-ND-N_{IJ}D_\mu X^{I}D^\mu\bar{X}^J+\frac12N_{IJ}Y^{Iij}Y_{ij}^J\nonumber\\
&+N_{IJ}f_{KL}^I\bar{X}^KX_Lf_{MN}^J\bar{X}^MX^N+\left(-\frac{\rm i}{4}\bar{F}_{IJ}F^{+I}_{\mu\nu}F^{+\mu\nu J}+{\rm h.c.}\right)\Biggr],
\end{align}
where $F_{IJ}(X)=\partial_I\partial_J F(X)$ and $F(X)$ is the prepotential with $w=2$, $N_{IJ}=-{\rm i}F_{IJ}+{\rm i}\bar{F}_{IJ}=2{\rm Im}F_{IJ}$, and $N=N_{IJ}X^I\bar{X}^J$. We rewrite the action in terms of the $N=1$ chiral and gauge multiplet components, $(X^I,P_L\chi^I, {\bf F}^I)$, $(A^I_\mu,\lambda^I,{\bf D}^I)$ and $(B_\mu, \hat\chi, \hat{D})$:
\begin{align}
S_{\rm vec}|_B=\int d^4x e\Biggl[&-\frac{1}{6}NR+\frac43N\hat{D}-N_{IJ}D_\mu X^{I}D^\mu\bar{X}^J+N_{IJ}\bar{\bf F}^I{\bf F}^J+\frac{1}{2}({\rm Im}F_{IJ}){\bf D}^I{\bf D}^J\nonumber\\
&-{\rm i}(N_{IL}\bar{X}^L)(X^Kf_{KJ}^I){\bf D}^{J}-\frac{1}{4}({\rm Im}F_{IJ})F^{I}_{\mu\nu}F^{\mu\nu J}-\frac{\rm i}{4}({\rm Re}F_{IJ})F^I_{\mu\nu}\tilde{F}^{J\mu\nu}\Biggr],\label{vectorS}
\end{align}
where we have used the property $f_{K(I}^LN_{J)L}=0$. We find that the $N=2$ vector multiplet action is written as that of $N=1$ conformal supergravity with 
\begin{equation}
\mathcal N= N(X,\bar{X}),\qquad W=0, \qquad f_{AB}= -{\rm i}F_{IJ}.
\end{equation}
We note that the second term on the first line of \eqref{vectorS} originates from the $U(1)_B$ charge of $X^I$, $n_B=\frac{4}{3}$, or correspondingly the Killing vector, $k_B^I=\frac{4\rm i}{3}X^I$. As we will see, the gauge multiplet of $B_\mu$ does not have kinetic terms either in the vector or the hypermultiplet action. Therefore, it behaves as an auxiliary superfield, which can be integrated out in obtaining physical action.

Next, let us discuss the hypermultiplet action, particularly its bosonic part given by
\begin{align}
S_{\rm hyper}|_B=\int d^4xe\left[-\frac{1}{12}k_D^2R+\frac14k_D^2D-\frac12g_{XY}D_\mu q^XD^\mu q^Y-2\bar{X}^IX^Jk_I^Xk_{JX}+P_{Iij}Y^{Iij}\right],\label{hyper}
\end{align}
where $k_D^2=k_D^Xg_{XY}k_D^Y$,
\begin{equation}
g_{XY}=(f^{i\bar{\cal A}}{}_X)^*d^{\bar{\cal A}}_{\cal B}f^{i\cal B}{}_Y=f^{i\cal A}{}_X\varepsilon_{ij}C_{\cal AB}f^{j\cal B}{}_Y,
\end{equation}
and 
\begin{equation}
P_{Iij}=\varepsilon_{k(i}A^{k\cal A}f^Y{}_{j)\cal A}k_I^Xg_{XY}=\varepsilon_{k(i}\varepsilon_{j)l}A^{k\cal A}C_{\cal AB}k_I^X\nabla_XA^{l\cal B}.
\end{equation}
$C_{\cal AB}$ is a covariantly constant anti-symmetric tensor, which is defined by $C_{\cal AB}=\rho_{{\cal A}\bar{\cal C}}d^{\bar{\cal C}}_{\cal B}$.
The square of Killing vector $k_{D}^2$ can be rewritten as
\begin{equation}
k_D^2=-2k_D^Xf^{1\cal A}{}_XC_{\cal BA}f^{2\cal B}{}_Yk_D^Y=2\Phi^{\cal A}d^{\bar{\cal B}}_{\cal A}\bar\Phi_{\bar{\cal B}}.\label{kdsq}
\end{equation}
We show the components of $P_{Iij}$ explicitly,
\begin{align}
P_{I11}=&-A^{2\cal A}C_{\cal AB}k_I^X\nabla_XA^{2\cal B}=-\bar{t}_I{}^{\bar{\cal A}\bar{\cal B}}\bar{\Phi}_{\bar{\cal A}}\bar{\Phi}_{\bar{\cal B}},\nonumber\\
P_{I22}=&-A^{1\cal A}C_{\cal AB}k_I^X\nabla_XA^{1\cal B}=-t_{I\cal AB}\Phi^{\cal A}\Phi^{\cal B},\nonumber\\
P_{I12}=&(A^{1\cal A}C_{\cal AB}k^I_X\nabla_XA^{2\cal B}+A^{2\cal A}C_{\cal AB}k^X_I\nabla A^{1\cal B})=2\bar{\Phi}_{\bar{\cal B}}d^{\bar{\cal B}}_{\cal A}\Phi^{C}t_{I\cal CD}C^{\cal AD},\label{PY}
\end{align}
where we have assumed the gauge invariance of $k_D^2$.
Using \eqref{kdsq} and \eqref{PY}, we rewrite the action~\eqref{hyper} as
\begin{align}
S_{\rm hyper}|_B=&\int d^4xe\Biggl[-\frac{1}{6}\Phi^{\cal A}d^{\bar{\cal B}}_{\cal A}\bar\Phi_{\bar{\cal B}}R-\frac23 \Phi^{\cal A}d^{\bar{\cal B}}_{\cal A}\bar\Phi_{\bar{\cal B}}\hat D-\nabla_\mu\Phi^{\cal A}d_{\cal A}^{\bar{\cal B}}\nabla^\mu\bar{\Phi}_{\bar{\cal B}}-4\bar{X}^I\bar{t}_I^{\bar{\cal B}\bar{\cal C}}\bar{\Phi}_{\bar{\cal B}}(d^{-1})^{\cal A}_{\bar{\cal C}}X^Jt_{J\cal AD}\Phi^{\cal D}\nonumber\\
&+\left(-\Phi^{\cal A}t_{I\cal AB}\Phi^{\cal B}{\bf F}^I+{\rm h.c.}\right)-{\rm i}\bar{\Phi}_{\bar{\cal B}}d^{\bar{\cal B}}_{\cal A}\Phi^{\cal C}t_{I\cal C}{}^{\cal A}{\bf D}^I+2\bar{\Phi}_{\bar{\cal B}}d^{\bar{\cal B}}_{\cal A}\Phi^{\cal C}t_{I\cal C}{}^{\cal A}f_{JK}^IX^J\bar{X}^{\bar K}\Biggr],\label{hypern1}
\end{align}

Except the last term, this action corresponds to the standard $N=1$ action with 
\begin{equation}
{\cal N}= \Phi^{\cal A}d^{\bar{\cal B}}_{\cal A}\bar\Phi_{\bar{\cal B}},\qquad  W=t_{I\cal AB}\Phi^{\cal A}\Phi^{\cal B}X^I.\label{HNW}
\end{equation}

We note that the second term in \eqref{hypern1} is consistent because $k_B^{\cal A}=-\frac{2\rm i}{3}\Phi^{\cal A}$. Also, our consideration about the on-shell F-term of $\Phi^{\cal A}$ discussed around \eqref{onshellF} is consistent with the superpotential~\eqref{HNW} and the K\"ahler potential $\cal N$. The forth term in \eqref{hypern1} can be regarded as the on-shell $F$-term scalar potential, $-{\bf F}^{\cal A}d_{\cal A}^{\bar {\cal B}}\bar{\bf F}_{\bar{\cal B}}$.

The last term on the second line of~\eqref{hypern1}, however, cannot be a part of the $N=1$ superconformal action.\footnote{One may think the shift of $Y_{12}^I$, which gives the extra term in \eqref{hypern1}, might be wrong. However, if this was the case, the vector action could not be consistent with the $N=1$ superconformal action formula. Therefore, either hyper- or vector multiplet action becomes inconsistent with $N=1$ superconformal formula. (Also, the transformation of vector multiplets could be inconsistent if we did not shift the definition of $Y_{12}^I$.)} Indeed, one can confirm that the extra term is necessary to reproduce the on-shell $N=2$ action after integrating out the auxiliary fields ${\bf D}$ and ${\bf F}^I$. The non-standard term cancels one term in the $N=1$ D-term potential, which is absent in $N=2$ supergravity potential.  We require the last term to vanish for consistency with $N=1$ superconformal symmetry,
\begin{equation}
P_{I12}f_{JK}^IX^J\bar{X}^{\bar K}=0.\label{cond3}
\end{equation}
 We will find that this condition is realized as the consistency with the parent $N=2$ theory, as we will discuss in Sec.~\ref{auxconst}.

In summary, we find that, under the truncation conditions~\eqref{psi2},\eqref{psi2sp} (and \eqref{cond3}), the $N=2$ superconformal action of the vector-hypermultiplet system can be reduced to the $N=1$ conformal supergravity with
\begin{equation}
{\cal N}=N(X,\bar{X})+\Phi^{\cal A}d_{\cal A}^{\bar{\cal B}}\bar{\Phi}_{\bar{\cal B}}, \qquad W=t_{I\cal AB}\Phi^{\cal A}\Phi^{\cal B}X^I,\qquad f_{AB}= -{\rm i}F_{IJ}(X).
\end{equation}
We also note that the chiral multiplet of $X^I$ and $\Phi^{\cal A}$  have $U(1)_B$ charges $n_B=4/3$ and $n_B=-2/3$, respectively. The $U(1)_B$ gauge multiplet does not have kinetic terms, that is, it is an auxiliary superfield. Integrating out the gauge multiplet $(B_\mu,\hat\chi, \hat D)$ gives constraints on other superfields as in the standard $N=2$ conformal supergravity.

Our reduced $N=1$ action is not yet consistent with the ``parent $N=2$'' system as the conditions~\eqref{psi2} and \eqref{psi2sp} can generally be inconsistent with their equations of motion. In order for the action to be fully consistent with the original $N=2$ supergravity, we have to check the equations of motion of the gravitino multiplet, particularly, the terms independent of the gravitino multiplet components. We will discuss such conditions in the next section.

\subsection{Consistency from the on-shell auxiliary fields}\label{auxconst}
We have shown that under the conditions~\eqref{psi2}, \eqref{psi2sp} and \eqref{cond3}, the $N=2$ superconformal action of vector and hyper-multiplets is described in terms of the standard $N=1$ conformal supergravity. So far, we have just imposed the truncation conditions~\eqref{psi2} and \eqref{psi2sp}, and have not yet discussed whether such conditions are consistent with their equations of motions. In particular, the terms linear in gravitino multiplet components $(\psi_\mu^2, V_{\mu}{}_1{}^2, T_{ab}^-, \chi^2)$ appear in the $N=2$ Lagrangian, which give the terms generally non-vanishing in their equations of motion. In order for such terms to vanish, more constraints on physical multiplets need to be imposed as we will discuss below.

 Let us look at the constraints from the auxiliary fields $V_\mu{}_1{}^2$ and $T_{ab}^+$. Under the truncation conditions \eqref{psi2} and \eqref{psi2sp}, following is required from their equations of motion,
\begin{align}
0=&\frac14N_{IJ}X^I\hat{F}^{+J}_{ab}+\frac16N_{IJ}X^I\bar{\Omega}^{2J}\gamma_a\psi^1_b-\frac{\rm i}{16}\bar{F}_{IJK}\bar{\Omega}^{1I}\gamma_{ab}\Omega^{2J}X^K\nonumber\\
&-\frac{\rm i}{12}d^{\bar{\cal A}}_{\cal B}A^{2\cal B}\bar{\zeta}_{\bar{\cal A}}\gamma_a\psi^1_b,\label{Tcond}\\
0=&2A^{2\cal A}C_{\cal AB}f^{2\cal B}_Y\tilde{D}_\mu q^Y-{\rm i}\bar\zeta_{\bar{\cal A}}\gamma^a\gamma^bA^{2\cal B}d_{\cal B}^{\bar{\cal A}}\psi_{a1}-\frac12N_{IJ}\bar{\Omega}^{2I}\gamma_\mu\Omega_{1}^J\nonumber\\
&-\frac{\rm i}{3}d^{\bar{\cal A}}_{\cal B}A^{2\cal B}\bar\zeta_{\bar{\cal A}}\gamma_{\mu\nu}\psi^\nu_1-\frac12\bar\psi_a^1\gamma^{ab}{}_\mu\psi_{b1}A^{2\cal A}C_{\cal AB}A^{2\cal B},\label{Vcond}
\end{align}
where $\tilde{D}_\mu q^X=D_\mu q^X|_{V_\mu{}_1{}^2=V_\mu{}_2{}^1}=0$.
The constraint from the equation of motion of the Lagrange multiplier $\chi^2$ under~\eqref{psi2} is
\begin{equation}
N_{IJ}X^I\Omega^{2J}-2{\rm i}d^{\bar{\cal A}}_{\cal B}A^{2\cal B}\zeta_{\bar{\cal A}}=0.\label{chiconst}
\end{equation}

For instance,
\begin{equation}
\bar{\Phi}_{\bar{\cal A}}C^{\bar{\cal A}\bar{\cal B}}\bar{\Phi}_{\bar{\cal B}}=0, \qquad N_{IJ}X^I\lambda^{J}=0,
\end{equation}
are required for consistency.
We note that the matrix $C^{\bar{\cal A}\bar{\cal B}}$ can be taken as
\begin{align}
C^{\bar{\cal A}\bar{\cal B}}=\left(
\begin{array}{cc}
0&\eta_{n_H}\\
-\eta_{n_H} &0 \\
\end{array} 
\right),\label{Cpara}
\end{align}
where $\eta_{n_H}$ is $n_H\times n_H$ matrix given by
\begin{align}
\eta_{n_H}=\left(
\begin{array}{cc}
{\mathbbm 1} _p&0\\
0&-{\mathbbm 1}_q \\
\end{array} 
\right)
\end{align}
and $p+q=n_H$.
This is realized by taking $\rho_{{\cal A}\bar{\cal B}}$ and $d^{\bar{\cal B}}_{\cal A}$ as
\begin{align}
\rho_{{\cal A}\bar{\cal B}}=\left(
\begin{array}{cc}
0& \mathbbm{1}_{n_H}\\
-\mathbbm{1}_{n_H} &0 \\
\end{array} 
\right),
\end{align}
\begin{align}
d^{\bar{\cal B}}_{\cal A}=\left(
\begin{array}{cc}
\eta_{n_H}& 0\\
0 &\eta_{n_H}\\
\end{array} 
\right).
\end{align}
 We separate the flat index into two parts, ${\cal A}=({\bf a},\dot{\bf a})$ where ${\bf a},\dot{\bf{a}}=1,\cdots,n_H$. Then, the condition $\bar{\Phi}_{\bar{\cal A}}C^{\bar{\cal A}\bar{\cal B}}\bar{\Phi}_{\bar{\cal B}}=0$ becomes
\begin{equation}
\bar{\Phi}_{\bar{\bf a}}\eta^{\bar{{\bf a}}\bar{\dot{\bf a}}}\bar{\Phi}_{\bar{\dot{\bf a}}}=0.\label{condp}
\end{equation}
We find that
\begin{equation}
\bar{\Phi}_{\bar{\dot{\bf a}}}=0
\end{equation}
is a nontrivial solution for the condition~\eqref{condp}, which implies the supersymmetric condition $\Phi^{\dot{\bf a}}(x, \theta)=0$. This condition removes the half of the chiral multiplets and solves all constraints for hypermultiplets implied by \eqref{Tcond}, \eqref{Vcond}, \eqref{chiconst}.

Next we consider the conditions on the $N=2$ vector multiplets. We impose
\begin{equation}
A^I_\mu=\lambda^I={\bf D}^I=0,
\end{equation}
for $I=a=1,\cdots, n_c$, and
\begin{equation}
X^I=P_L\chi^I={\bf F}^I=0
\end{equation} 
for $I=\alpha=n_c+1,\cdots,n_V$, where $n_c$ and $n_V$ denote the number of the $N=1$ chiral multiplet and the total number of the $N=2$ vector multiplet. Note that the gauge transformation of $X^{\alpha}=0$ reads $\delta X^\alpha=\theta^Kf_{bK}^{\alpha}=0$, namely, $f_{bc}^{\alpha}=f^\alpha_{b\beta}=0$ where $\theta^K$ is gauge transformation parameters. Similarly, from $\lambda^a=0$, $f^{a}_{\alpha \beta}=f^{a}_{\alpha b}=0$.  

The condition $N_{IJ}X^I\lambda^{J}=0$ implies $N_{a\alpha}X^a\lambda^{\alpha}=0$ and also the third term on the right-hand-side of \eqref{Tcond} vanishes if $\bar{F}_{ab\alpha}X^b=0$. Therefore, we require
\begin{equation}
\frac{\partial}{\partial X^{a}}\frac{\partial}{\partial X^\alpha}N=\frac{\partial}{\partial X^a}\frac{\partial}{\partial \bar{X}^\alpha}N=0,
\end{equation}
or equivalently,
\begin{equation}
N_{a\alpha}=\bar{F}_{ab\alpha}X^b=0.
\end{equation}
One can show that $T_{ab}^+=V_{\mu}{}_1{}^2=0$ as well as the constraint from the equation of motion of $\chi^2$~\eqref{chiconst} can be solved by the conditions
\begin{equation}
\bar{\Phi}_{\bar{\dot{\bf a}}}=0(=\Phi^{\dot{\bf a}}),\qquad N_{a\alpha}=\bar{F}_{ab\alpha}X^b=0.\label{const4}
\end{equation}

Finally, we discuss the condition that the terms linear in $\psi^2_\mu$ vanish. Under the condition~\eqref{const4}, we find the following terms linear in $\psi^2_\mu$ and $\psi_{\mu2}$,
\begin{align}
{\cal L}_{\psi^2}=&\Bigg(\frac{\rm i}{2}\bar{F}_{a\alpha}\bar{\Omega}^{1a}\gamma_{\mu}\psi^2_\nu\tilde{\hat{F}}^{\mu\nu\alpha}-\frac{\rm i}{2}\bar{F}_{a\alpha}\bar{X}^a\psi^1_\mu\psi^2_\nu\hat{F}^{\mu\nu\alpha}+\frac12N_{ab}\bar{\Omega}^a_1f^b_{cd}\gamma^\mu\psi_{\mu 2}X^c\bar{X}^d+2{\rm i}\bar{X}^a\bar{\zeta}_{\bar{\bf a}}\gamma^{\mu}\psi^2_{\mu}d^{\bar{\bf a}}_{\bf b}t_{a\bf c}{}^{\bf b}\Phi^{\bf c}\nonumber\\
&+\frac12\bar{\psi}_{\mu 2}\gamma^{\mu}\Omega_1^aP_{a}^{12}+\frac12\bar{\psi}_{\mu2}\gamma^{\mu}\Omega_{2}^\alpha P_{\alpha 22}+\frac12\bar{X}^a\bar{\psi}^1_{\mu}\gamma^{\mu\nu}\psi^2_{\nu}P_{a12}+{\rm h.c.}\Bigg)\, .\label{plinear}
\end{align}
The first two terms imply the condition stronger than \eqref{const4}, $\bar{F}_{a\alpha}=0.$
The third term vanishes if $f_{bc}^{a}=0$. The second term in the second line of \eqref{plinear} vanishes if $P_{\alpha22}=0$, which implies $t_{\alpha\bf ab}=0$. $P_{a12}=0$ is realized by the condition $t_{a\bf a}{}^{\bf b}=0$ that also removes the fourth term in the first line of \eqref{plinear}. Thus, we find a set of the consistency conditions,
\begin{equation}
\Phi^{\dot{\bf a}}=0,\quad \bar{F}_{a\alpha}=0,\quad f_{bc}^a=0,\quad t_{\alpha \bf ab}=0=t_{a\bf a}{}^{\bf b}.\label{consistency}
\end{equation}
Note that $t_{a\bf a}{}^{\bf b}=0$ implies $t_{a{\bf a}\dot{\bf b}}=0$ since we use $C_{\cal AB}$ for lowering of the index (see the parametrization \eqref{Cpara}). Also, only the nonzero structure constant should be $f_{JK}^I=f_{\beta\gamma}^\alpha$. The set of consistency conditions lead to \eqref{cond3}, and therefore we need no more requirements.

In summary, we find that the consistent $N=1$ truncation of $N=2$ conformal supergravity is described by
\begin{equation}
S=[N(X^a,\bar{X}^a)+\Phi^{\bf a}d_{\bf a}^{\bar{\bf b}}\bar{\Phi}_{\bar{\bf b}}]_D+[\tilde{t}_{a\bf ab}\Phi^{\bf a}\Phi^{\bf b}X^a]_F+[-{\rm i}F_{\alpha\beta}(X^a)\bar\lambda^\alpha P_L\lambda^\beta]_F,\label{final}
\end{equation}
where $N=X^a\bar{X}^b(-{\rm i}F_{ab}+{\rm i}\bar{F}_{ab})$. The truncation conditions are \eqref{psi2} and \eqref{psi2sp}, and their consistency conditions with the $N=2$ action are \eqref{consistency}. We note that similar consistency conditions are found in \cite{Andrianopoli:2001zh,Andrianopoli:2001gm}

\subsection{Comments on superconformal gauge fixing}
We give some comments about the $N=1$ superconformal gauge fixing conditions. It has been known that the $N=2$ conformal supergravity with vector and hypermultiplets requires one vector and one hyper compensator multiplet. Correspondingly, in our reduced $N=1$ system, we need two chiral compensators, one from $X^a$ and another from $\Phi^{\bf a}$, whose kinetic term has negative sign.\footnote{The constraints~\eqref{const4} are also important for reducing the number of negative norm multiplets. For example, if we did not truncate half of the chiral multiplets $\Phi^{\cal A}$, we would have two chiral compensators with negative kinetic terms, and one of them would be left as a physical ghost.} The reason is the following: The equation of motion of the auxiliary field $\hat D$ reads
\begin{equation}
\frac 43N-\frac23 \Phi^{\bf a}d^{\bar{\bf b}}_{\bf a}\bar{\Phi}_{\bar{\bf b}}=0,
\end{equation}
and the dilatation gauge fixing condition which makes graviton canonical is
\begin{equation}
N+ \Phi^{\bf a}d^{\bar{\bf b}}_{\bf a}\bar{\Phi}_{\bar{\bf b}}=-3,
\end{equation}
in the Planck unit, $M_{\rm pl}=1$. These conditions lead to
\begin{equation}
N=X^{a}N_{ab}\bar{X}^{b}=-1,\qquad \Phi^{\bf a}d^{\bar{\bf b}}_{\bf a}\bar{\Phi}_{\bar{\bf b}}=-2.
\end{equation}
Therefore, for each set of chiral multiplets $X^a$ and $\Phi^{\bf a}$, there should be a multiplet with a negative definite metric to solve these conditions. The $U(1)_B$ charge difference between $X^a$ and $\Phi^{\bf a}$ is crucial for the gauge fixing procedure, and this is why we need both hyper- and gauge multiplet compensator in $N=2$ conformal supergravity.

In our $N=1$ language, the elimination of two chiral compensators are understood as follows: One of them is removed by the standard superconformal gauge fixing. The other one is eaten by the auxiliary gauge multiplet of $B_\mu$, and the massive auxiliary gauge multiplet is integrated out after all. Thus we find only the physical multiplets with positive kinetic terms. 

Once we derive the standard $N=1$ conformal supergravity system, the superconformal gauge fixing procedure is the same as the standard one (see e.g. \cite{Freedman:2012zz}). We do not repeat it here. Note that, one may integrate out the $U(1)_B$ gauge multiplet after superconformal gauge fixing since they are commutative.

\section{Summary}\label{summary}
We have discussed the consistent reduction of off-shell $N=2\to N=1$ conformal supergravity. As we have shown, the full $N=2$ theory has $N=1$ subgroup, under which the $N=2$ Weyl multiplet is represented by the $N=1$ Weyl, the gravitino, and U(1) gauge multiplet. Truncating the gravitino multiplet~\eqref{psi2}, \eqref{psi2sp} consistently realizes the standard rules of $N=1$ conformal supergravity. As we have shown explicitly, $N=2$ vector and hyper-multiplets simply become $N=1$ vector and chiral pairs and two chiral pairs, respectively. Keeping off-shell structure is useful to see the agreement with the standard $N=1$ transformation laws. We have found that the $N=2$ superconformal action under truncation condition can be described by the standard $N=1$ rules, except one term~\eqref{hypern1}. We also discussed the consistency of the truncation condition, and it turned out that the consistency with the original $N=2$ theory requires to truncate half of the $N=1$ multiplets. We have derived relatively simple truncation conditions \eqref{consistency}, and the resultant $N=1$ conformal action is given by the standard formulae with \eqref{final}.
 
We comment on possible extensions of this work. In this work, we did not study the truncation of the other multiplets such as a tensor multiplet, and similar analysis would be possible for such multiplets as in Poincar\'e supergravity analysis~\cite{DAuria:2005ipx}. Also, the reduction of $N=4$ conformal supergravity would also be possible in a similar way. A possible application of our off-shell formalism would be to describe the reduction of $N=2$ models with higher-derivative terms, where the consistent reduction would be more difficult for on-shell Lagrangian. Constructing phenomenological and cosmological model building would also be an interesting direction. We will study these possibilities for future work.
\section*{Acknowledgement}
The author is grateful to Hiroyuki Abe, Shuntaro Aoki, Daniel Butter for discussions on related issues and Renata Kallosh for comments on the manuscript.
The author is supported by Stanford Institute for Theoretical Physics and by the US National Science Foundation Grant  PHY-1720397.
\appendix
\section{$N=2$ superconformal algebra and transformation laws of multiplets}\label{A}
$N=2$ superconformal group consists of the set of generators $(P_a, Q^i_\alpha, S_{i\alpha}, M_{ab}, D, U_i{}^j, T, K_a)$, where $P_a$ is translation, $Q^i$ supersymmetry, $S_i$ $S$-supersymmetry, $M_{ab}$ Lorentz rotation, $D$ dilatation, $U_i{}^j$ SU(2) R-symmetry, $T$ chiral U(1), and $K_a$ special conformal boost. The index $a$ is for a local Lorentz index, $i$ for SU(2), and $\alpha$ for spinor index.  We have a set of gauge fields $(e_\mu^a,\psi^i_\mu, \phi_{i\mu}, \omega^{ab}_{\mu},b_\mu, V_i{}^j{}_\mu, A_\mu, f_{\mu}^a)$ for each generator. We use the notation that $Q^i$ and $S_i$ are left handed, and accordingly, $\psi^i_\mu$ and $\phi_{i\mu}$ are left handed, $\psi^i_\mu=P_L\psi_\mu^i$. Complex conjugation raises or lowers SU(2) indices. The nontrivial commutators of $N=2$ superconformal algebra are as follows.
\begin{align}
&\{Q_\alpha^i,Q_j^\beta\}=-\frac12\delta_j^i(\gamma^a)_{\alpha}^\beta P_a,\quad \{S_\alpha^i, S^\beta_j\}=-\frac{1}{2}\delta_j^i(\gamma^a)_\alpha^\beta K_a,\quad \nonumber\\
&\{Q_{i\alpha},S^{j\beta}\}=-\frac12\delta_i^j\delta_\alpha^\beta D-\frac14\delta_i^j(\gamma^{ab})_\alpha^\beta M_{ab}+\frac{\rm i}{2}\delta_i^j\delta_\alpha^\beta T-\delta_\alpha^\beta U_{i}{}^j,\nonumber\\
&\{Q_{\alpha}^i,S^{\beta}_j\}=-\frac12\delta^i_j\delta_\alpha^\beta D-\frac14\delta^i_j(\gamma^{ab})_\alpha^\beta M_{ab}-\frac{\rm i}{2}\delta^i_j\delta_\alpha^\beta T-\delta_\alpha^\beta U_{j}{}^i,\nonumber\\
&[D,P_a]=P_a,\quad [D,K_a]=-K_a,\quad [D,Q_\alpha^i]=\frac12Q_\alpha^i,\quad [D,S_\alpha^i]=-\frac12S_\alpha^i,\nonumber\\
&[M_{ab},M_{cd}]=-2\eta_{c[a}M_{b]d}+2\eta_{d[a}M_{b]c},\nonumber\\
&[M_{ab},P_c]=P_a\eta_{bc}-P_b\eta_{ac},\quad [M_{ab},Q_{\alpha}^i ]=-\frac12(\gamma_{ab}Q^i)_\alpha,\nonumber\\
&[M_{ab},K_c]=K_a\eta_{bc}-K_b\eta_{ac},\quad [M_{ab},S_{\alpha}^i ]=-\frac12(\gamma_{ab}S^i)_\alpha,\nonumber\\
&[T,Q_\alpha^i]=\frac{\rm i}{2}Q^i_\alpha,\quad [T,S_\alpha^i]=\frac{\rm i}{2}S_\alpha^i, \quad [U_i{}^j,Q_\alpha^k]=\delta_i^kQ_\alpha^j-\frac12\delta_i^jQ^k_\alpha,\nonumber\\
&[U_i{}^j,S_\alpha^k]=\delta_i^kS_\alpha^j-\frac12\delta_i^jS^k_\alpha, \quad [U_i{}^j,Q_{\alpha k}]=-\delta_k^jQ_{\alpha i}+\frac12\delta_i^jQ_{\alpha k},\nonumber\\
&[U_i{}^j,S_{\alpha k}]=-\delta_k^jS_{\alpha i}+\frac12\delta_i^jS_{\alpha k},\quad [U_i{}^j,U_k{}^l]=\delta_i^lU_k{}^j-\delta_k^jU_i{}^l,\nonumber\\
&[K_a,Q_\alpha^i](\gamma_a S^i)_\alpha, \quad [P_a,S_\alpha^i]=(\gamma_a Q^i)_\alpha.\label{N2algebra}
\end{align}

We also show the $Q$- and $S$-transformations of the independent gauge fields and auxiliary fields:
\begin{align}
\delta e_\mu^a=&\frac12\bar\epsilon^i\gamma^a\psi_\mu+{\rm h.c.},\\
\delta \psi_\mu^i=&\left(\partial_\mu+\frac12b_\mu+\frac14\gamma^{ab}\omega_{\mu ab}-\frac{\rm i}{2}A_\mu\right)\epsilon^i+V_\mu{}^i{}_j\epsilon_j-\frac{1}{16}\gamma^{ab}T_{ab}^-\varepsilon^{ij}\gamma_\mu\epsilon_j-\gamma_\mu\eta^i,\\
\delta b_\mu=&\frac12\bar\epsilon^i\phi_{\mu i}-\frac12\bar\eta^i\psi_{\mu i}-\frac38 \bar\epsilon^i\gamma_\mu \chi_i+{\rm h.c.},\\
\delta A_\mu=&-\frac{\rm i}{2}\bar\epsilon^i\phi_{\mu i}-\frac{\rm i}{2}\bar\eta^i\psi_{\mu i}-\frac{3\rm i}{8}\bar\epsilon^i\gamma_\mu\chi_i+{\rm h.c.},\\
\delta V_{\mu}{}_i{}^j=&-\bar{\epsilon}_i\phi^j_\mu-\bar\eta_i\psi^j_\mu+\frac34\bar\epsilon_i\gamma_\mu\chi^j+\bar\epsilon^j\phi_{\mu i}+\bar\eta^j\psi_{\mu i}-\frac34\bar\epsilon^j\gamma_\mu\chi_i\nonumber\\
&-\frac12\delta_i^j(-\bar\epsilon_k\phi^k_\mu-\bar\eta_k\psi^k_\mu+\frac34\bar\epsilon_k\gamma_\mu\chi^k+\bar\epsilon^k\phi_{\mu k}-\frac34\bar\epsilon^k\gamma_\mu\chi_k),\\
\delta T^-_{ab}=&2\bar\epsilon^i\hat R_{ab}(Q^i)\varepsilon_{ij},\\
\delta \chi^i=&\frac12 D\epsilon^i +\frac16\gamma^{ab}\left[-\frac14\slashed{D}T^-_{ab}\varepsilon^{ij}\epsilon_j-\hat{R}_{ab}(U_j{}^i)\epsilon^j+{\rm i}\hat{R}_{ab}(T)\epsilon^i+\frac12 T^-_{ab}\varepsilon^{ij}\eta_j\right],\\
\delta D=&\frac12\bar\epsilon^i \slashed{D}\chi_i+{\rm h.c.},
\end{align} 
where $T^{-}_{ab}$ is an anti-self dual tensor, $\chi^i$ is left-handed spinor, and $D$ is a real scalar. We use $D_\mu$, through out this paper, as the covariant derivative under superconformal and internal gauge symmetries. We also show the curvatures which appear in the transformations,
\begin{align}
\hat R_{\mu\nu}(Q^i)=&2\left(\partial_{[\mu}+\frac12b_{[\mu}+\frac14\gamma_{ab}\omega^{ab}_{[\mu}-\frac{\rm i}{2}A_{[\mu}\right)\psi^i_{\nu]}+2V_{[\mu}{}^i{}_j\psi_{\nu]}^j-\frac18\gamma^{ab}T_{ab}^-\varepsilon^{ij}\gamma_{[\mu}\psi_{\nu] j}-2\gamma_{[\mu}\phi^i_{\nu]},\\
\hat R_{\mu\nu}(U_i{}^j)=&2\hat{\partial}_{[\mu}V_{\nu]}{}_i{}^j+\left(-4\bar\psi_{[\nu i}\phi^j_{\mu]}-4\bar\phi_{[\nu i}\psi^j_{\mu]}+\frac32\bar\psi_{[\nu i}\gamma_{\mu]}\chi^j-\frac32\bar\psi^j_{[\nu}\gamma_{\mu]}\chi_i\right)\Bigg|_{\rm traceless},\\
\hat{R}_{\mu\nu}(T)=&2\partial_{[\mu}A_{\nu]}+\left(-{\rm i}\bar\psi^i_{[\nu}\phi_{\mu]i}-{\rm i}\bar\phi^i_{[\nu}\psi_{\mu]i}-\frac{3\rm i}{4}\bar\psi^i_{[\nu}\gamma_{\mu]}\chi_{i}+{\rm h.c.}\right),
\end{align}
where $A_{[\mu} B_{\nu]}=\frac12 (A_\mu B_\nu-A_\nu B_\mu)$, $\hat\partial_\mu V_{\nu}{}_i{}^j$ denotes the covariant derivative with respect to SU(2), and $A_i^{j}|_{\rm traceless}$ is the projection to make $A_i{}^j$ traceless.

The matter representations, vector and hypermultiplets consist of $(X^I, \Omega^I_i, A_\mu^I, Y_{ij})$ and $(q^X, \zeta^{\cal A})$, respectively. Their $Q$- and $S$-transformation laws are as follows: For a gauge multiplet,
\begin{align}
\delta X^I=&\frac12\bar\epsilon^i\Omega_i^I,\\
\delta\Omega^I_i=&\slashed{D}X^I\epsilon_i+\frac14\gamma^{ab}{\cal F}_{ab}^I\varepsilon_{ij}\epsilon^j+Y_{ij}^I\epsilon^j+X^I\bar{X}^Kf_{JK}^I\varepsilon_{ij}\epsilon^j+2X^I\eta_i,\\
\delta A^I_{\mu}=&\frac12\varepsilon^{ij}\bar\epsilon_i\gamma_\mu\Omega_j^I+\varepsilon^{ij}\bar\epsilon_i\psi_{\mu j}X^I+{\rm h.c.},\\
\delta Y_{ij}=&\frac12\bar\epsilon_{(i}\slashed{D}\Omega_{j)}^I-f_{JK}^I\bar\epsilon_{(i}\varepsilon_{j)k}\Omega^{kK}X^J-\frac12\varepsilon_{k(i}\varepsilon_{j)l}\bar\epsilon^k\slashed{D}\Omega^{Il}-f_{JK}^I\bar\epsilon^k\varepsilon_{k(i}\Omega^{K}_{j)}\bar X^J,
\end{align}
where
\begin{equation}
{\cal F}^{I}_{\mu\nu}\equiv F^{I}_{\mu\nu}-\left(\varepsilon_{ij}\bar\psi^i_{[\mu}\gamma_{\nu]}\Omega^{Ij}+\varepsilon_{ij}\bar\psi^i_\mu\psi_\nu^j\bar{X}^I+\frac12\bar{X}^IT^-_{ab}+{\rm h.c.}\right),
\end{equation}
and $F_{\mu\nu}^I=2\partial_{[\mu}A_{\nu]}^I+A_{\mu}^JA_{\nu}^Kf^I_{JK}$.

For a hypermultiplet,
\begin{align}
\delta q^X=&-{\rm i}\bar\epsilon^i\zeta^{\cal A}f^X{}_{i\cal A}+{\rm i}\varepsilon^{ij}\rho^{\bar{\cal A}{\cal B}}\bar\epsilon_i\zeta_{\bar{\cal A}}f^X{}_{j\cal B},\\
\delta \zeta^{\cal A}=&\frac{\rm i}{2}f^{i\cal A}{}_X\slashed{D}q^X\epsilon_{i}-\zeta^{\cal B}\omega_{X\cal B}{}^{\cal A}\delta q^X+{\rm i}\bar{X}^Ik_I^Xf^{i\cal A}{}_X\varepsilon_{ij}\epsilon^j.
\end{align}
More details are shown in Sec.~\ref{hyperreduction}.

\section{$N=1$ superconformal symmetry and multiplets}\label{B}
We show the $N=1$ superconformal algebra and $Q$- and $S$-transformation laws of the Weyl, a gauge and a chiral multiplet. The algebra consists of $(P_a, Q_\alpha, M_{ab},D,A, S_\alpha, K_a)$. The commutation relations are
\begin{align}
&\{Q_\alpha,Q^\beta\}=-\frac12(\gamma^a)_{\alpha}^\beta P_a,\quad \{S_\alpha, S^\beta\}=-\frac{1}{2}(\gamma^a)_\alpha^\beta K_a,\quad \nonumber\\
&\{Q_{\alpha},S^{\beta}\}=-\frac12\delta_\alpha^\beta D-\frac14(\gamma^{ab})_\alpha^\beta M_{ab}+\frac{\rm i}{2}(\gamma_*)_\alpha^\beta A,\nonumber\\
&[D,P_a]=P_a,\quad [D,K_a]=-K_a,\quad [D,Q_\alpha]=\frac12Q_\alpha,\quad [D,S_\alpha]=-\frac12S_\alpha,\nonumber\\
&[M_{ab},M_{cd}]=-2\eta_{c[a}M_{b]d}+2\eta_{d[a}M_{b]c},\quad [M_{ab},P_c]=P_a\eta_{bc}-P_b\eta_{ac},\quad [M_{ab},Q_{\alpha} ]=-\frac12(\gamma_{ab}Q)_\alpha,\nonumber\\
&[M_{ab},K_c]=K_a\eta_{bc}-K_b\eta_{ac},\quad [M_{ab},S_{\alpha} ]=-\frac12(\gamma_{ab}S)_\alpha,\quad [A,Q_\alpha]=-\frac{3\rm i}{2}(\gamma_*Q)_\alpha,\nonumber\\
&[T,S_\alpha]=\frac{3\rm i}{2}(\gamma_*S)_\alpha, \quad [K_a,Q_\alpha^i](\gamma_a S^i)_\alpha, \quad [P_a,S_\alpha^i]=(\gamma_a Q^i)_\alpha.\label{N1algebra}
\end{align}

The Weyl multiplet has four independent fields $(e^a_\mu, \psi_\mu, b_\mu, \hat{A}_\mu)$ and others are dependent. The $Q$- and $S$-transformation laws of independent fields are as follows:
\begin{align}
\delta e^a_\mu=&\frac12\bar\epsilon\gamma^a\psi_\mu,\\
\delta \psi_\mu=&\left(\partial_\mu+\frac12 b_\mu+\frac14\omega_\mu^{ab}\gamma_{ab}-\frac{3\rm i}{2}A_\mu\gamma_*\right)\epsilon-\gamma_\mu\eta,\\
\delta b_\mu=&\frac12\bar\epsilon\phi_\mu-\frac12\bar\eta\psi_\mu,\\
\delta \hat{A}_\mu=&-\frac{\rm i}{2}\bar \epsilon\gamma_*\phi_\mu+\frac{\rm i}{2}\bar\eta\gamma_*\psi_\mu.
\end{align}

A vector multiplet $(A_\mu^A, \lambda^A, D^A)$ has a vector $A_\mu^A$, Majorana spinor $\lambda^A$ and a real scalar $D^A$, whose transformation laws are
\begin{align}
\delta A_\mu^A=&-\frac12\bar\epsilon\gamma_\mu\lambda^A,\label{N1gauge}\\
\delta\lambda^A=&\left(\frac14\gamma^{ab}\hat{F}_{ab}^A+\frac{\rm i}{2}\gamma_*D^A\right)\epsilon\label{N1gaugino},\\
\delta D^A=&\frac{\rm i}{2}\bar\epsilon\gamma_*\slashed{D}\lambda^A,\label{N1D}
\end{align}
where $\hat F^A_{\mu\nu}=2\partial_{[\mu}A_{\nu]}^A+f_{BC}^AA_\mu^BA_\nu^C+\bar\psi_{[\mu}\gamma_{\nu]}\lambda^A$ and $f_{BC}^A$ is a structure constant. All components are $S$-inert.

A chiral supermultiplet $(\Phi^I, P_L\chi^I, F^I)$ made of a complex scalar $\Phi^I$, a left-handed Weyl spinor $P_L\chi^I$ and a complex auxiliary field $F^I$ has the following transformation laws:
\begin{align}
\delta \Phi^I=&\frac{1}{\sqrt2}\bar\epsilon P_L\chi^I,\label{N1CS}\\
\delta P_L\chi=&\frac{1}{\sqrt2}P_L(\slashed{D}\Phi^I+F^I)\epsilon+\sqrt2 w\Phi^IP_L\eta,\label{N1CW}\\
\delta F^I=&\frac{1}{\sqrt2}\bar\epsilon\slashed{D}P_L\chi+\sqrt{2}(1-w)\bar\eta P_L\chi+\bar\epsilon k^I_AP_R\lambda^A,\label{N1F}
\end{align}
where $w$ is the Weyl weight corresponding to the charge under dilatation $D$. We have assumed that the multiplet is gauged and $k^I_A$ is the Killing vector $\delta X^I=\theta^Ak_{A}^I$ and $\lambda^A$ denotes a gaugino. Note that a chiral multiplet satisfies $w=n$ where $n$ is the charge under $A$,called the chiral weight. The above expression is not covariant under the coordinate transformation of the scalar manifold spanned by $X^I$. The covariant formulation of chiral multiplets can be found in \cite{Freedman:2016qnq,Freedman:2017obq}.

\bibliographystyle{JHEP}
\bibliography{ref}
\end{document}